\documentclass[preprint,5p,authoryear]{elsarticle}

\usepackage[final]{microtype}
\usepackage{xcolor}
\usepackage{harpoon}
\usepackage{amssymb}
\usepackage{textcomp}
\usepackage{gensymb}
\usepackage{hyperref}
\usepackage{graphicx}
\usepackage{wrapfig}
\usepackage{rotating}
\usepackage{caption}
\usepackage{subcaption}
\usepackage{hhline}
\usepackage{chngpage}
\usepackage{float}
\usepackage{adjustbox}
\usepackage{amsmath}

\setbool{@fleqn}{false}

\graphicspath{ {./images/} }

\journal{Astronomy and Computing}

\begin{document}
\begin{frontmatter}

\title{\texttt{DUG Insight}: A software package for big-data analysis and visualisation, and its demonstration for passive radar space situational awareness using radio telescopes}

\author[inst1,inst2]{Grigg, D.}
\ead{dylang@dug.com}

\affiliation[inst1]{organization={DUG Technology},
            addressline={76 Kings Park Rd}, 
            city={West Perth},
            postcode={6005}, 
            state={WA},
            country={Australia}}

\author[inst2]{Tingay, S.J.}
\author[inst2]{Sokolowski, M.}
\author[inst2]{Wayth, R.B.}

\affiliation[inst2]{organization={International Centre for Radio Astronomy Research, Curtin University},
            addressline={Kent St}, 
            city={Bentley},
            postcode={6102}, 
            state={WA},
            country={Australia}}

\begin{abstract}

As the demand for software to support the processing and analysis of massive radio astronomy datasets increases in the era of the SKA, we demonstrate the interactive workflow building, data mining, processing, and visualisation capabilities of \texttt{DUG Insight}. We test the performance and flexibility of \texttt{DUG Insight} by processing almost 68,000 full sky radio images produced from the Engineering Development Array (EDA2) over the course of a three day period. The goal of the processing was to passively detect and identify known Resident Space Objects (RSOs: satellites and debris in orbit) and investigate how radio interferometry could be used to passively monitor aircraft traffic. These signals are observable due to both terrestrial FM radio signals reflected back to Earth and out-of-band transmission from RSOs. This surveillance of the low Earth orbit and airspace environment is useful as a contribution to space situational awareness and aircraft tracking technology. From the observations, we made 40 detections of 19 unique RSOs within a range of 1,500 km from the EDA2. This is a significant improvement on a previously published study of the same dataset and showcases the flexible features of \texttt{DUG Insight} that allow the processing of complex datasets at scale. Future enhancements of our \texttt{DUG Insight} workflow will aim to realise real-time acquisition, detect unknown RSOs, and continue to process data from SKA-relevant facilities.

\end{abstract}


\begin{keyword}

Space situational awareness \sep \texttt{DUG Insight} \sep radio astronomy \sep interactive workflow creation \sep high performance computing
\end{keyword}

\end{frontmatter}

\section{INTRODUCTION}
\label{sec:sample1}

Radio astronomy has consistently been at the forefront of the data challenge in science. Using interferometry, large arrays of antennas can be formed and the data streams from all antennas can be combined, processed and time averaged, reducing output data volumes \citep{book}. Aggregate data rates have evolved from MB/s to GB/s in the current generation of interferometric radio telescopes. Examples include the Murchison Widefield Array (MWA) in Western Australia \citep{2013PASA...30....7T,2018PASA...35...33W} and the LOw Frequency ARray (LOFAR) mostly in the Netherlands \citep{2013A&A...556A...2V}. These radio interferometers are precursors and pathfinders for the SKA - planned to be the largest radio interferometer ever constructed - for which the output data rates will reach TB/s \citep{2020SPIE11449E..0XC}.

To meet this future data challenge, significant effort is being placed into the data processing software and computational platforms for the SKA. For example, an imaging pipeline that processed full-scale simulated SKA data on the Summit supercomputer generated 2.6 PB of output data \citep{2020hpcn.conf...11W}. Beyond the core data processing, these massive output datasets require further processing in order for SKA users to extract astrophysical information from them. This will be achieved via a global network of SKA Regional Centres (SRCs) \citep{2020SPIE11449E..0XC}.

SKA users will require efficient methods to manipulate datasets on the petabyte to potentially exabyte scale. In order to undertake quality control, assess the complex choices made in imaging and calibration processing, and to extract astrophysical information, work is being undertaken globally to develop software for these tasks. For example, CARTA (Cube Analysis and Rendering Tool for Astronomy) is an analysis tool designed for image visualisation and analysis at the scale of SKA precursor telescopes, with the aim of being scalable for future telescopes \citep{2021zndo...4905459C}. Large-scale data transfer between MeerKAT, an SKA precursor in South Africa \citep{2012AfrSk..16..101B} and the UK, has been explored to utilise processing capacity at the IRIS (Incorporated Research Institutions for Seismology) computing facilities \citep{2021arXiv210514613T}.

In this paper we describe the \texttt{DUG Insight} software package\footnote{https://dug.com/dug-insight/} (henceforth referred to as \texttt{Insight}). \texttt{Insight} was developed to support the analysis of seismic data for oil and gas exploration, where the data exist in multi-dimensional volumes. \texttt{Insight} is designed to handle massive datasets, with the ability to utilise large-scale back-end compute and storage. Workflows for data manipulation and processing can utilise built-in standard signal processing methods, as well as other bespoke processing functions, or scripts supplied by the user. The results of workflows can be visualised in real-time, allowing highly interactive processing for the user. \texttt{Insight} is described in more detail in Section \ref{sec:insight}.

\begin{figure}
    \centering
    \includegraphics[width=0.46\textwidth]{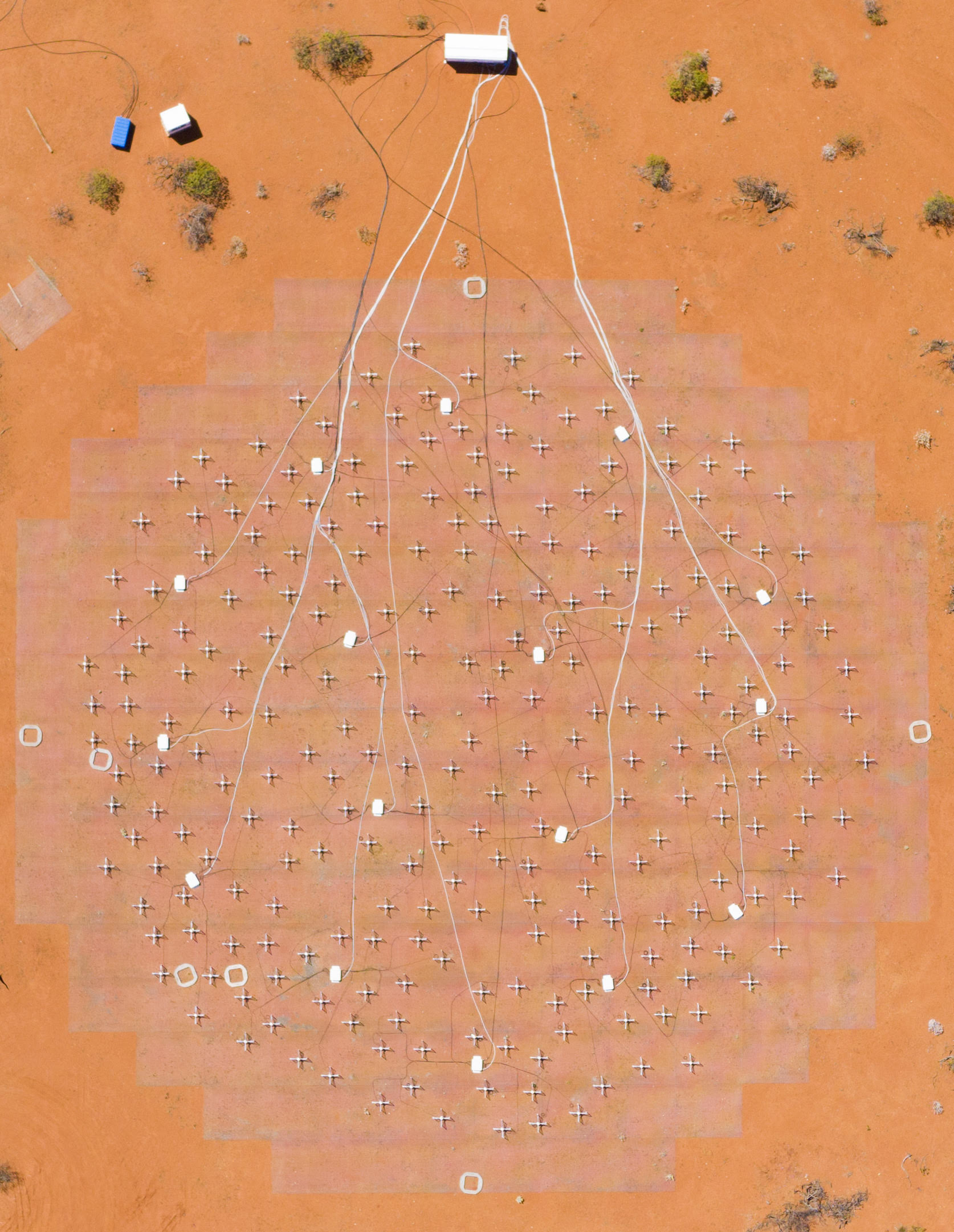}
    \caption{\label{fig:eda2} The Engineering Development Array 2, the instrument used for this work.}
\end{figure}

We demonstrate the application of \texttt{Insight} to radio astronomy data. The data structures, data volumes, and signal processing algorithms common in radio astronomy are similar to those encountered in seismic data processing. While the dataset we have chosen for this demonstration is not of large volume, it is complex and contains signals from astrophysical and anthropogenic radio sources. The data are from an interferometric array that is being used to benchmark the performance of SKA antenna technologies, the so-called Engineering Development Array 2 (EDA2; Figure \ref{fig:eda2}) \citep{2021arXiv211200908W}. The array consists of 256 individual low frequency antennas arranged on a 35 m diameter footprint, representing the architecture of a single SKA `station'. The Western Australian based low frequency SKA will consist of 512 similar stations spread over thousands of square kilometres.

The images produced by the EDA2 cover the entire sky and have been generated at high cadence over an extended observational period (every $\sim$8 s over 75 hours). The radio frequency used for the observations has been purposefully chosen to lie within the FM broadcast band and tuned to the frequency of an FM transmitter $>$300 km away. Characterising the effect of transmitters in this frequency range for radio telescopes at the site of the low frequency SKA was a goal of this research.

The dataset used here has been previously analysed for these purposes using a Python-based bespoke analysis code, which revealed a range of interesting signals \citep{2020PASA...37...39T}. In addition to astrophysical radio emission across the sky, the analysis found signals received directly from FM transmitters, as well as the same FM signals reflected off aircraft, meteor ionisation trails, and objects in Earth orbit known as Resident Space Objects (RSOs) - satellites and space junk.

This last category of signal is interesting, as the increasing population of RSOs is beginning to impact Earth-based astronomy \citep{2001.10952}, to the point where papers such as \citet{space_environmentalism} are drafted in an effort to deter courts from allowing the mass launches of satellites. The International Astronomical Union says it is ``Deeply concerned about the increasing number of launched and planned satellite constellations in mainly low Earth orbits'' and have formed the Centre for the Protection of the Dark and Quiet Sky from Satellite Constellation Interference\footnote{\url{https://www.iau.org/science/scientific_bodies/centres/CPS/}} to address this issue.

While the human-made signals are regarded as noise (Radio Frequency Interference: RFI) in most areas of astronomy, this information is useful for Space Situational Awareness (SSA). The detection and tracking of RSOs can be used to assist in managing the space environment. In this context, the array used here is an example of a passive bi-static radar system, opportunistically utilising FM transmitters \citep{2020PASA...37...52P}.

As an interesting signal class with astronomical and space science significance, we focus our \texttt{Insight} demonstration on improving the detection and characterisation of FM transmissions reflected off RSOs, relative to the previously presented rudimentary initial analysis of these signals \citep{2020PASA...37...39T}. A goal of this research is to use publicly available information for known RSOs to guide our detection algorithms. We also aim to more comprehensively survey the airspace for aircraft to determine how they affected the dataset and our results. The search for these complex signals provides a comprehensive demonstration of \texttt{Insight}'s capabilities.

We also aim to complement the work of \citet{soko_eda2} which uses the EDA2 and the Aperture Array Verification System 2 (AAVS2; \citep{10.1117/12.2562391}), a similar 256 antenna radio interferometer, to characterise astrophysical radio transients in the Southern Hemisphere.

In Section \ref{sec:insight}, we present an overview of the \texttt{Insight} package. In Section \ref{sec:obs}, we describe the observational data and our data processing using \texttt{Insight}. In Section \ref{sec:results} we present our results, Section \ref{sec:discussion} details the discussion, and conclusions follow in Section \ref{sec:conclusion}.

\section{\texttt{Insight}}
\label{sec:insight}

\texttt{Insight} is an interactive software package for seismic data quality control, qualitative/quantitative interpretation, visualisation, workflow development, and multidimensional big data-analysis. With custom technology for different types of input data, \texttt{Insight} allows advanced signal processing and imaging workflows to minimise the risk associated with hydrocarbon exploration. 

A knowledge of High Performance Computing (HPC) and scripting is not necessary in the regular usage of \texttt{Insight}, allowing the user to be more research and results focused. The back-end algorithms connecting the software to the HPC interface allow the use of a wide range of modern architectures best suited to the job. Streamlined tools allow the rapid iteration of investigation and data manipulation with interactive visualisation. Lengthy command-line compilation stages or the need to orchestrate disparate tools and scripts is avoided. Custom standalone, non-interactive processes can also be run through the \texttt{Insight} framework to have the job submission and output handling managed by the software and in a GUI.

Fundamental, highly optimised geophysical algorithm's full waveform inversion \citep{10.1093/gji/ggt258} and  reverse time migration \citep{ZHOU2018207} have been run on 170,000 cores for $\sim$447,000 node hours and $\sim$421,000 simultaneous cores, respectively. This demonstrates both the scalability of \texttt{Insight} and the stability of the HPC system the software is coupled with.

To demonstrate the applicability of \texttt{Insight} to other domains such as radio astronomy, the following general example highlights how a typical \texttt{Insight} project functions. 

\texttt{Insight} features a comprehensive, customisable data-loading tool to deal with a diverse range of formats which may not be standardised. If necessary, the raw data are ingested through this tool into \texttt{Insight}'s optimised data format. This stores adjacent data in a bundle which it refers to as a `brick', selected based on the data type to maximise throughput and minimise unnecessary I/O. This allows efficient interactive processing and visualisation without sacrificing speed in production runs or ease of use at scale.

Quality control (QC) of the dataset generally follows and \texttt{Insight} hosts an array of tools to interrogate the raw data for anomalies. QC of multi-terabyte datasets needs to be performed quickly but comprehensively and the QC tools give the user the ability to perform global QC that identifies anomalies that can be analysed in fine detail. Examples are acquisition related errors, such as a faulty sensor, or oversights by the technical team, such as poor calibration, or unavoidable events, such as lightning, which can either render sections of the data unusable or necessitate additional filtering. One example of these tools is the RMS map, which is designed to give a broad overview of the quality of the dataset.

Dynamic processing workflows are then created to manipulate data according to user defined rubrics. Through this system, the user can interactively test their data processing choices. The workflow is structured so that each processing step represents a filter or operation with an interactive output that can be visualised to help achieve optimal parameterisation. The user can arbitrarily filter or rearrange the dimensionality of their datasets for both visualisation and processing. For example, this could be utilised to view a section of every image in the dataset simultaneously to identify local anomalies. This expedites the time to achieve optimal parameterisation and prevents the user from spending time submitting/waiting for a job to finish, which is inherent to a HPC dependent workflow system. 

The user can generate or manually apply a variety of annotation tools to highlight locations or events in the dataset. These annotations can then be integrated as parameters in the processing workflow. A variety of visualisations such as histograms, graphs, and two/three dimensional representations of the data can become increasingly relevant as processing progresses. The visualisation of attributes across the data (typically in five or more dimensions) allows validation of the processing quality.

For advanced users who wish to customise their own processes and filters, \texttt{Insight} features an interactive Python Integrated Development Environment (IDE) to cater for custom algorithms. The IDE allows the user to interactively test their script live within the processing workflow, with changes in the script being reflected in the visualisations of the output. This allows for rapid debugging and optimisation of Python code through interactive execution on intermediate process outputs. For users who wish to use other more performant languages, there is functionality to plug straight in to Insight's own architecture (Java based) which allows for connection to native and external applications through the JNI (Java Native Interface) for executing C or C\texttt{++} code.

\begin{figure*}[ht]
    \centering
    {\includegraphics[width=\textwidth]{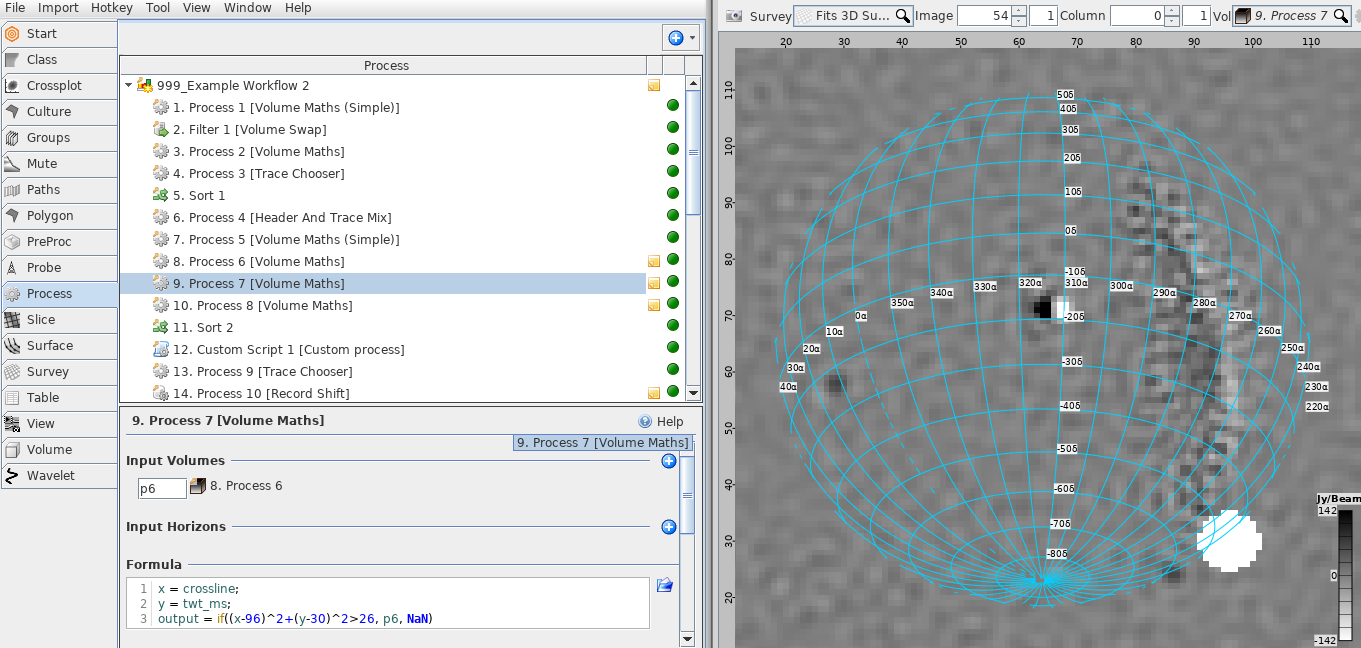}}
    \caption{\label{fig:gui} A typical view of the \texttt{Insight} Graphical User Interface.}
\end{figure*}

\begin{figure*}[t]
    \centering
    {\includegraphics[width=1\textwidth]{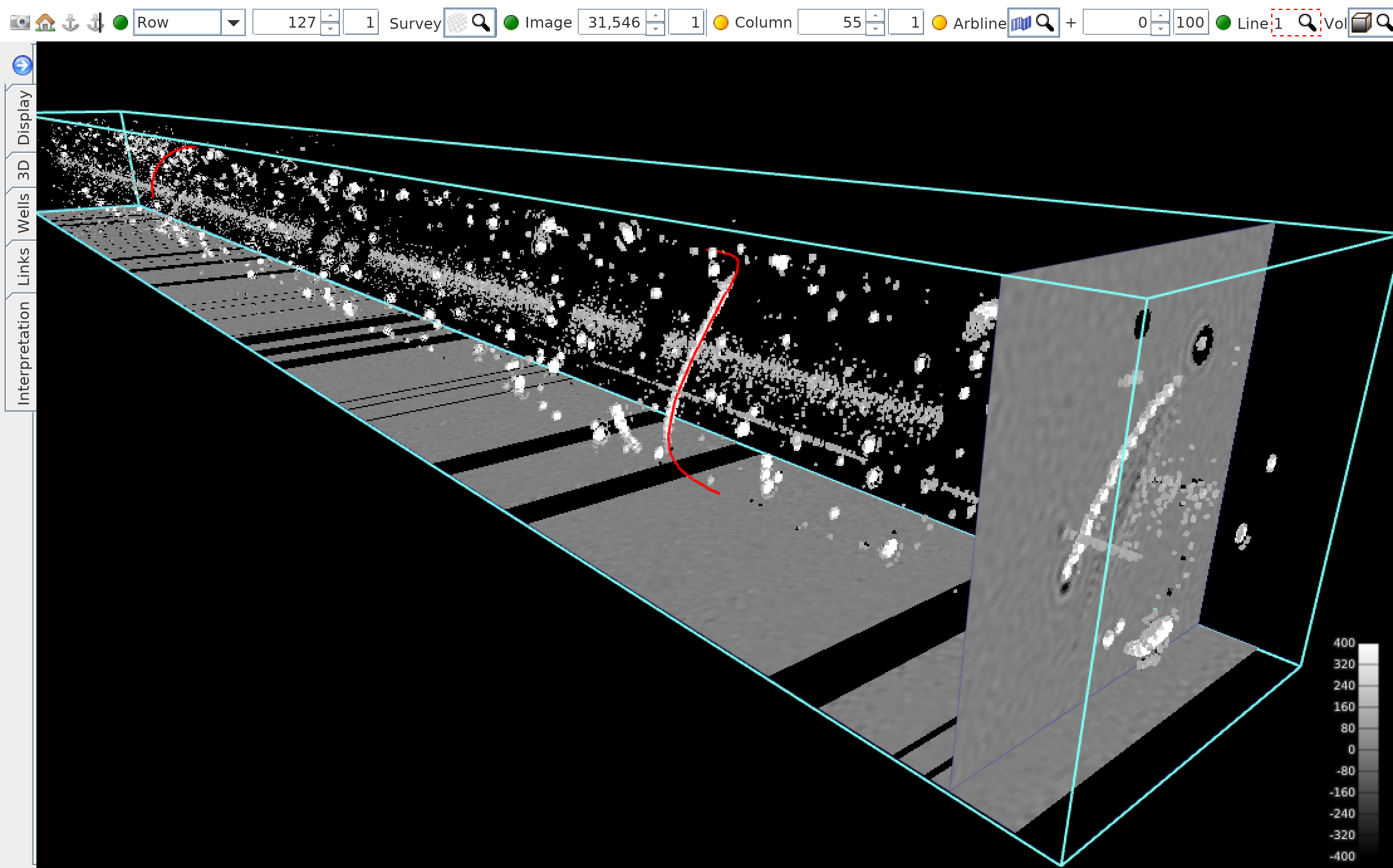}}
    \caption{\label{fig:3d_viewer} The \texttt{Insight} 3D viewer. Pictured are the difference images of the EDA2 dataset described in this paper as a semi-transparent pseudo-3D volume. The image with index 31,546 is rendered in full on the right hand side, as well as the bottom row of pixels for each image along the base of the cyan prism. The gaps are due to noisy images being removed prior to this processing step. All data displayed have passed through 12 steps of processing which has been computed interactively in this visualisation. The red curves are annotations derived from external data on the predicted trajectory of the International Space Station (NORAD ID: 25,544).}
\end{figure*}

Figures \ref{fig:gui} and \ref{fig:3d_viewer} show screenshots of the Graphical User Interface of the \texttt{Insight} software. Figure \ref{fig:gui} shows the typical working setup of the software. On the left is the control panel where the user performs the processing, selects items to view, submits jobs to the HPC cluster, and controls the software. An example workflow is pictured with a custom code process that produces the circular white cutout on the bottom right hand side of the image. The interactive output of `Process 7' from the workflow is being visualised in the image viewer on the right. A 3D spherical projection of a generalised orthographic world coordinate system has been superimposed over the image from coordinate information provided by the dataset's metadata. This allows exploration of the dataset from varying angles in multiple dimensions. 

The 3D viewer can be seen in Figure \ref{fig:3d_viewer}. Here a 2D dataset is being viewed as a pseudo-3D volume where each image occupies one plane of voxels. Increasing the image index is equivalent to moving forward in time. In this example, the lower negative amplitudes have been made transparent and this accentuates internal features of the volume. There is a pattern of detected signals over the red predicted trajectory.

The dataset can be sliced in various dimensions (such as viewing an image, viewing one column of pixels across multiple images, or viewing one row of pixels across multiple images) and viewed simultaneously across multiple interconnected visualisation windows at any granularity. This interconnection enables the user to flexibly interrogate their data and highlight trends or areas of interest for further processing. For this example, the input dataset to the workflow forms the only data on disk. The 12 steps of processing being visualised in the 3D viewer have been interactively computed. This way there is no waiting for jobs to finish and no unnecessary intermediate products are stored to disk as they are instead stored in memory for immediate analysis.

The software interfaces to the HPC system through an adapter to the cluster's queueing system. The benefit of running jobs through the Insight GUI are that outputs and workflow metadata are stored in a predictable location aiding in the reproducibility of results. It also provides a tool to give the user fine control of products on-disk through the GUI which is useful for large projects.

\texttt{Insight} is also available on an expanding list of cloud providers.

The general framework set out by \texttt{Insight} provides an efficient transition into other large-scale signal processing domains. The size of a production seismic dataset (or volume as it is known in \texttt{Insight}) is usually on the order of terabytes to petabytes. Such dataset sizes are now common in modern radio astronomy and a niche that \texttt{Insight} can fill is the on-the-fly interaction with these datasets in the transition to exascale processing. \texttt{Insight}'s QC tools have been identified as a potential point of interest to explore, for flagging poor quality data and removing artefacts. The visual interaction with data has been highly beneficial in seismic data processing, even at the petabyte scale. To exercise this point, the study described here aims to visualise, run QC analysis, and process a radio astronomy dataset. This is outlined in Section \ref{sec:obs}.

While the reader may have questions regarding whether Insight can be used as a comprehensive tool for their radio astronomy project, it is important to note that the software is constantly evolving. The trajectory of this development is to ensure that a user can complete their work efficiently and with scientific rigour so that results can be reported and reproduced for future interrogation. Part of this growth of the software comes from discerning users' needs to guide the development. Further discussion on this is always invited.

\section{OBSERVATIONS AND DATA PROCESSING}
\label{sec:obs}

The dataset used for this work was acquired using the Engineering Development Array 2 (EDA2) over three days spanning $31^{\rm st}$ January - $3^{\rm rd}$ February 2020. The observations and data processing are described in detail by \citet{2020PASA...37...39T}.

\begin{figure*}
    \centering
    \includegraphics[width=0.9\textwidth]{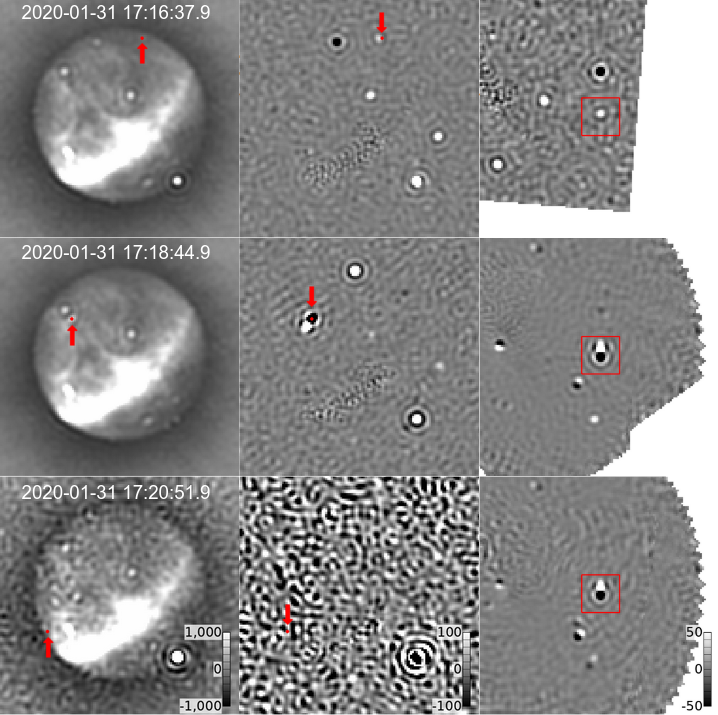}
    \caption{\label{fig:image_pane}Three timesteps of combined XX and YY polarisation images (left column), corresponding difference images (middle column), and cumulative stacks of the International Space Station (ISS; NORAD 25,544) after shifting and stacking (right column). In order from top to bottom for the cumulative stacks, they have fold numbers of 1, 17, and 33 which represents the number of difference images stacked together one-by-one over the course of the ISS pass. In the third timestep, the RMS noise in the image is higher due to a bright RFI burst from Geraldton in the bottom right corner which contaminates the image with noise. In the left and central columns, the predicted location of the ISS from the TLE information is annotated with a red dot with an arrow to guide the reader's eye. The red square in the right column images is to guide the reader's eye to the centred ISS position in the stacked images.}
\end{figure*}

Briefly, the EDA2 is composed of 256 dual-polarisation MWA dipole antennas in a 35 m diameter footprint. Used as an imaging interferometer, the calibrated and processed EDA2 data produced 33,944 images in each of the two polarisations (XX=East-West dipole; YY=North-South dipole) over the course of the three day period. Each image used 7.92 s of data produced in a 0.926 MHz bandwidth with a central frequency of 98.4375 MHz. The images cover the entire visible sky, horizon to horizon, from the EDA2 location (26.703056 S, 116.672268 E). Example images are displayed in the left column in Figure \ref{fig:image_pane}.

\begin{figure}
    \centering
    \includegraphics[width=0.46\textwidth]{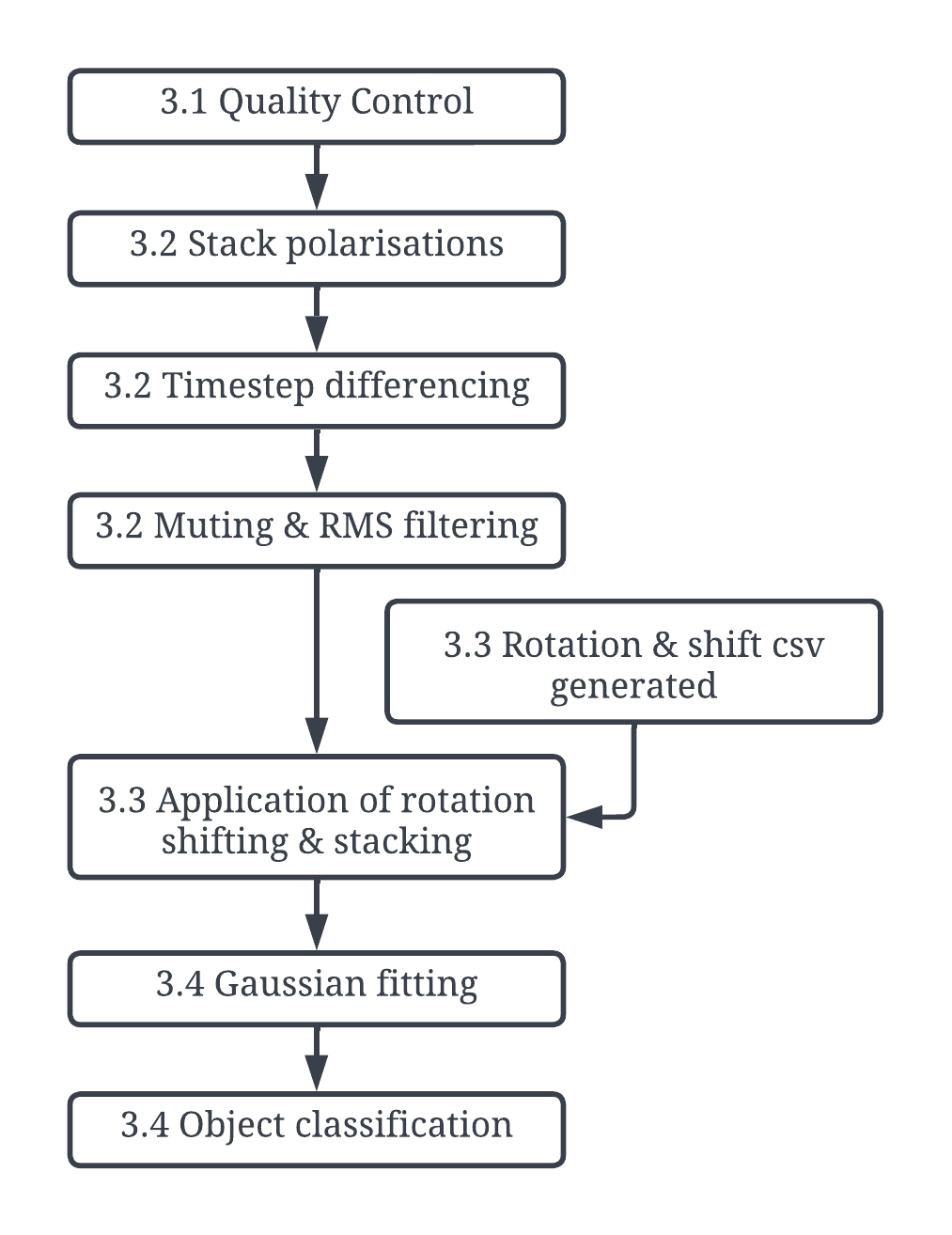}
    \caption{\label{fig:flowchart} A visual representation of the major processing steps.}
\end{figure}

The input data to the \texttt{Insight} processing were 33,994 XX and 33,994 YY FITS images which had been prepared with \texttt{MIRIAD} \citep{1995ASPC...77..433S}. This processing was also outlined in the previous work \citep{2020PASA...37...39T}. The customisable nature of the Insight data-loading tool allowed ingestion of the two polarisations separately into Insight's optimised data format.

The flowchart pictured in Figure \ref{fig:flowchart} shows a visual representation of the processing workflow explained in the remainder of this section.

\subsection{Data Integrity and Quality Control Checks}

\texttt{Insight}'s toolkit features an array of data integrity and QC tools which were used to identify both acquisition based anomalies (e.g. one of the EDA2 dipole sensors is faulty) or data dependent anomalies (e.g. RFI).

A pseudo 3D volume was visualised to identify any dataset level anomalies. The images were ordered sequentially in time and assigned to a voxel as can be seen in Figure \ref{fig:3d_viewer}. This included the ability to slice through the volume spatially and temporally in a cross-image sense to highlight unusable images. Images containing a high level of noise were immediately visually identifiable and could be scrutinised in more detail if needed. This gave an immediate understanding for the amount of RFI contamination in the dataset. The final 94 images from each polarisation were removed, as this QC highlighted that they contained pixel values that were consistently very low. Further investigation based on this analysis revealed that hardware failure rendered the images unusable.

To ensure no images were missing, the hour, minute, and second from the UTC of each image were plotted and analysed for anomalies. There were no breaks in the regular 7.92 s acquisition of each image. The coordinate systems described in the image metadata were separately plotted for each image, to ensure there were no coordinate system errors in the dataset.

\subsection{Data Preparation}
\label{sec:data_prep}

\begin{figure*}[ht]
    \centering
    \includegraphics[width=1.0\textwidth]{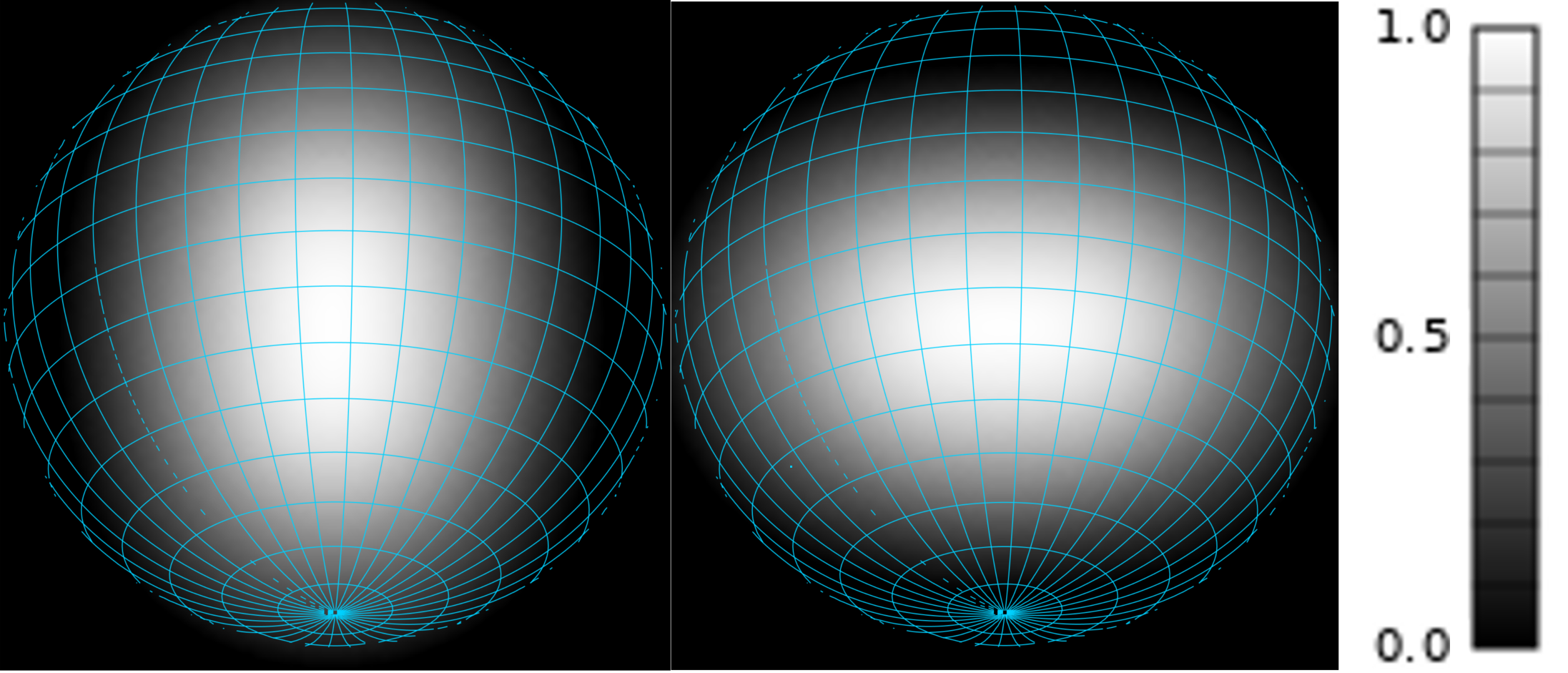}
    \caption{\label{fig:beams} The simulated beam power patterns of a single EDA2 dipole antenna for an infinite ground plane (X polarisation is left and Y polarisation is right). These have been normalised. The projection is included to show where the pattern ends at the horizon.}
\end{figure*}

The following processing was performed and parameterised interactively in \texttt{Insight} through the data transformation pipeline. The two polarisations were stacked to both increase the Signal to Noise Ratio (SNR) and to reduce the bias of the polarisation beam pattern which is shown in Figure \ref{fig:beams}. A simple average of the two polarisations produced the best result in terms of robustly increasing the SNR of detections and is defined as:

\begin{equation}\label{pol_stack}
I_{n}=\frac{XX_{n} + YY_{n}}{2}
\end{equation}

 where $I_{n}$ is the averaged image at time $n$, $XX_{n}$ is the XX polarisation image at time $n$, and likewise for $YY_{n}$.

Images at consecutive timesteps were then differenced ($\Delta I_{n}=I_{n+1}-I_{n}$). This substantially reduced pseudo-stationary (in time and/or position) sources of radio emission such as extragalactic point-like sources and the diffuse radio emission from the Milky Way. The central column in Figure \ref{fig:image_pane} shows an example of these difference images.

RFI signals from locations near Geraldton and directly south of the array (likely Perth, $\sim$800 km away) are highly visible in the data. A mute, which is the circular cutout parameterised in Figure \ref{fig:gui}, was applied over each of the towns so that this known source of RFI would not affect subsequent processing. This was parameterised and tested interactively in the \texttt{Insight} maths tool which allows the user to write a logical expression to manipulate the array values of the images. The formula section on the control panel in Figure \ref{fig:gui} shows how this was written and included as a process in the workflow.

Images which contained a source of high amplitude RFI would often be unusable due to sidelobe contamination across the image. Within a custom script, the RMS of each image was calculated as a process in the workflow. A cross-image display was used to view one column of pixels from each image with the RMS plotted above. This made interrogating a specific image with a high RMS straightforward and allowed an intuition to be built by testing different thresholds and deciding if images were correctly being removed. As a result of this testing, an RMS upper limit of 200 Jy/beam was found to optimally remove noisy images and retain usable images. 2,626 images were removed based on this criterion which is approximately 5.8 hours worth of observation time. The majority were removed due to very strong reflected signals from aircraft. Some specific instances of noisy images could also be removed manually when found in an interactive interrogation of the dataset.

\subsection{Rotation, Shifting and Stacking}
\label{sec:rotation}

To detect reflected terrestrial FM transmissions from RSOs close to or below the per-image noise level, images were stacked along the path of the RSO to increase its SNR. 

In difference space, the RSOs have a distinctive signature of two diffuse components: a positive feature followed by a negative feature \citep{2020PASA...37...52P} (henceforth referred to as the `difference signature'). The difference signature can be seen inside the red squares on the right hand column in Figure \ref{fig:image_pane}. It is a function of the beam response of the array and the motion of the RSO over the duration of the two observations ($\sim16$ seconds). The positive leading feature is the location of the RSO in frame $I_{n+1}$ and the negative feature is the location of the RSO in the previous frame, $I_{n}$. Along the RSO's trajectory, the difference signature will rotate as the RSO's position changes with respect to the array. To ensure the SNR is optimally increased when stacking, each image must be rotated and shifted to correct for this rotation and the movement of the RSO before stacking.

To calculate the rotation and stacking parameters, information is needed on the predicted orbital parameters of the RSOs. This information is recorded as a Two Line Element set (TLE) and is publicly available through the Space Surveillance Network\footnote{From \texttt{space-track.org}}. The TLE describes the orbital parameters of the RSO at an epoch from which the position of the object can be propagated for a particular time and observation location. Using the Skyfield Python module \citep{2019ascl.soft07024R} and the coordinate information contained in the FITS metadata, the predicted image pixel coordinates of the RSOs were calculated for each image. 

The position of each RSO within each image was calculated and this was made available to visualise in \texttt{Insight} as an annotation, such as the red curves in Figure \ref{fig:3d_viewer}. These annotations were used to confirm the identity of RSOs found in the dataset or to show the predicted position of an RSO when manually scrolling through images.

This study considered all RSOs within a perigee range of 400 - 1,500 km, also known as low Earth orbit. 6,110 RSOs matched this criterion between the dates 2020-01-31 to 2020-02-03. 

For a particular frame at time $t$, an RSO's predicted pixel location is propagated from the TLE and found to be (\textit{$x_{t}$}, \textit{$y_{t}$}). This point is translated to an origin which was arbitrarily chosen to be (64, 64) or point O - the centre of the 128$\times$128 pixel image. Any offset in the RSO's observed position from point O will be mainly due to inaccuracies in the TLE (generally due to the TLE being out of date). In this study the offsets were generally found to be less than one pixel.

\begin{figure}
    \centering
    \includegraphics[width=0.3\textwidth]{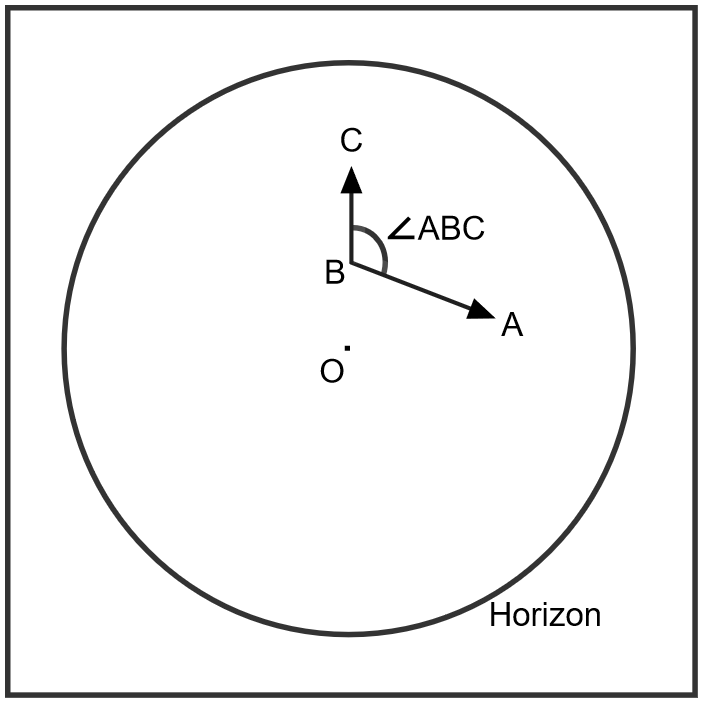}
    \caption{\label{fig:sat_schematic} A schematic to add visual clarity to the stacking procedure. The diagram is of an arbitrary image and the horizon is pictured for reference. The vectors are not to typical scale.}
\end{figure}

To rotate the difference signature correctly, (\textit{$x_{t+1}$}, \textit{$y_{t+1}$}), (\textit{$x_{t}$}, \textit{$y_{t}$}), and (\textit{$x_{t}$}, \textit{$y_{t} + 1$}) must be calculated, referred to as points A, B, and C, respectively and can be seen in Figure \ref{fig:sat_schematic}. The vector $\overrightharp{BA}$ represents the 2D distance and direction of travel of the RSO with respect to time $t$ and $\overrightharp{BC}$ represents the desired orientation of the RSO trajectory after rotation. This was chosen to orient the positive feature in the difference signature in the $+y$ direction and the negative feature in the $-y$ direction. To solve for the angle $\angle{ABC}$ (the angle needed to rotate the RSO) the following equation was used:

\begin{equation} \label{cosine_rule}
\angle{ABC} = \cos^{-1} \left({\frac{\overrightharp{BA} \cdot \overrightharp{BC}}{\vert \overrightharp{BA} \vert \vert \overrightharp{BC} \vert}}\right) .
\end{equation}

To correct $\angle{ABC}$ for the true clockwise angle, if $\textit{x}_{t+1} < \textit{x}_{t}$ then $\angle{ABC}$ needs to be subtracted from $360\degree$.

$\angle{ABC}$ is then calculated per image when an RSO is above the horizon and is applied with respect to the origin, for each RSO. Each time the RSO crosses the visible field of view, the images across the path are stacked together. For $N$ difference images in the pass, the stacked difference image ($\Delta I_{s}$) is:

\begin{equation} \label{stack}
\Delta I_{s}=\sum_{n=1}^{N} \Delta I_{n}
\end{equation}

There was no need to compensate for $N$ in this stack as the SNR and morphology classifications were performed on a per-stack basis meaning that normalisation had no effect on the results.

Through interactive testing of the \texttt{Insight} workflow, it was found that an RSO would generally only be detectable for the central third of the pass for the higher amplitude detections. This is where the RSO passes closest to the zenith and when the physical distance between the RSO and the EDA2 is at a minimum for that pass. Images outside this middle third only introduced noise into the stacking and were therefore not included in the stack.

As can be seen in the branch on the flowchart in Figure \ref{fig:flowchart}, the shifts and rotations were calculated once outside of \texttt{Insight} and this information was input into the workflow as a CSV file. Any of the 6,110 RSOs could then be instantly stacked for any pass, which was valuable for testing the workflow on specific RSOs. 

\subsection{Quantifying Detections}
\label{sec:quantifying}
Reproducibility and efficiency of the workflow was a major consideration, ensuring that meaningful results could be delivered at scale. Differentiating between detections and non-detections was crucial, as this workflow creates almost a hundred thousand stacks, most of which contain only noise. Therefore a critical consideration in the post processing analyses was how to quantify a detection.

The stacks were reduced to only contain a small sub-image region centred on the origin (point O). The sub-image extent was 9 x 14 pixels, totalling 126 pixels. The extents were chosen to include only data for which all rotated and shifted images contributed to the stack.

By interactively testing strong detections, it was found that some stacks with a small number of frames were produced when the RSO was only visible for a short period of time. This was often due to the RSO's trajectory remaining close to the horizon at a large angle to the zenith. There were no detections made with $N < 24$ in interactive or larger automated tests and therefore stacks which satisfied this condition were excluded. The RSO only being visible at large zenith angles means the RSO was further away from the array and therefore less likely to be detected.

To classify potential detections, a Gaussian function was fitted to each remaining stack to model the difference signature. Constraints on the morphology of the fitted Gaussians could then be imposed to systematically select detections. \texttt{Insight}'s Python IDE enabled the rapid development of this Gaussian fitting, allowing the code to be altered and tested interactively. This meant that no jobs had to be submitted to the HPC system and tests were only calculated locally on the machine as required. This was a particularly useful feature of \texttt{Insight} that increased the productivity of this project.

The Scipy \texttt{curve\_fit} algorithm \citep{2020SciPy-NMeth} was used to fit the Gaussian function to the difference signature, with the function consisting of two equal but inverted amplitude ($A$) Gaussians (one with a positive amplitude and one with a negative amplitude). These were separated on the y-axis of the stack by a distance $y_{0}=|y_{0,pos}|+|y_{0,neg}|$, where $y_{0,pos}$ and $y_{0,neg}$ are offsets from $y=0$ for the positive Gaussian and the negative Gaussian, respectively. Both Gaussians are offset from $x=0$ by $x_{0}$ and share a common value for $\sigma$. Expressed in full,
\begin{equation} \label{gaussian}
\begin{split}
I_{model}=A(&e^{-C((x - x_{0})^{2} + (y - y_{0,pos})^{2})} - \\&e^{-C((x - x_{0})^{2} + (y - y_{0,neg})^{2})}) ,
\end{split}
\end{equation}
where $C = \frac{1}{2 \sigma^{2}}$.

Thus, the fitted parameters are $A$, $x_{0}$, $y_{0,pos}$, $y_{0,neg}$, and $\sigma$ (via $C$). Starting values for these parameters were provided to the \texttt{curve\_fit} algorithm to initiate the fitting. These were based on the expected position of the RSO based on the rotation and shifting correction, the maximum and minimum values in the small sub-image region, and a sigma of 1. Providing reasonable starting values assisted in the convergence of the fit.

The quality of the fitting was tested interactively on strong detections by subtracting the fit from the stack to examine the residuals. Good fits result in noise-like residuals.

The SNR of detections was defined as the fitted value of $A$ (an estimate of the signal amplitude) divided by the square root of the covariance of the fitted value of $A$ (an estimate of the error on the fitted parameter). We adopt a detection threshold of ${\rm SNR} > 6$ and only examine stacks that meet this criterion. 

We also select physically plausible morphologies for the difference signature. We only select stacks that have a fitted value of $\sigma<2$, which corresponds to a full width at half maximum (FWHM) for the Gaussian of $2\sigma \sqrt{2 ln(2)} \sim 2.355\sigma \sim 5$ pixels. This is chosen because the FWHM of the EDA2 resolution element is approximately five pixels in extent (calculated as $\sim \lambda/D$, where $\lambda$ is the wavelength of the radio waves and $D$ is the diameter of the EDA2). Difference signatures with fitted sizes greater than five pixels cannot be RSOs.

Similarly, the separation between the two Gaussians cannot be more than a maximum physically plausible distance, set by the fastest and closest RSOs producing the maximum possible angular speed and therefore separation between two images eight seconds apart. We have set the upper and lower limits for the Gaussian separation to be $2.5 < |y_{0,pos}|+|y_{0,neg}| < 7$ pixels which equate to altitudes of $\sim400$ km and $\sim1,100$ km respectively. Note we have included TLEs up to a perigee of 1,500 km as these separations are an approximation. We expect noise to affect the appearance of the separation in some instances and the diffuse nature of the difference signature to decrease the accuracy of separation measurements. 

Interactive analysis showed that the expected position of an RSO generated from TLE information was generally never more than one pixel away from the observed position in the horizontal direction and never more than two pixels in the vertical direction. This extra variability in the vertical axis is likely due to contraction of the difference signature from different observed angular speeds of the RSOs at each time step.

As a final criterion, to prevent poor fits, the absolute value of the flux density at the maximum ($x_{0}, y_{0,pos}$) and minimum ($x_{0}, y_{0,neg}$) of the fitted model could not be more than twice the input stack values at the same locations. This criterion discards examples where the algorithm produces poor fits due to either a highly dominant positive or negative Gaussian component.
 
The stacks which met these criteria were deemed candidate RSO detections and were subjected to further analysis. The flexible nature of \texttt{Insight}'s internal format allowed the output dataset to be ordered so that the individual difference images could be placed next to each stack for quick and efficient further interrogation of each candidate detection. The final analysis was performed on this volume.

Once the parameterisation of the workflow was deemed optimal, it was submitted as a job through the \texttt{Insight} UI so that the output could be written to disk. Running this job in \texttt{Insight} has multiple benefits. No Slurm, HPC cluster management software, or command line interaction was required. Especially important for large scale projects is the full logging of parameters and job execution for reproducibility of results, with the output stored in a predictable directory in the project structure.

For $N$ stacked images with a constant signal and Gaussian noise, the noise is $\propto 1/\sqrt{N}$. In this workflow, there are two passes of stacking which were described in Equations \ref{pol_stack} and \ref{stack} and as discussed, the minimum value of $N$ allowed was 24. With each image being the stack of two polarisations, a minimum SNR improvement of $\sqrt{48}$ can be expected compared to a single image at a single polarisation. This is an approximation, as in reality the noise is not completely Gaussian in nature and the amplitude of the source may vary.

An additional test was run to determine the probability of false positives. The processing flow described above was re-run with the image indices shifted forwards by 10 and was limited to 10,000 RSO flyovers. This meant that the predicted TLE trajectories did not track RSOs. The results of this test aided in discerning the likelihood of the stack containing a false positive detection. It also gave an indication of the false positive probability for each category of detection listed in Section \ref{sec:results}, as no reflected terrestrial radio energy was expected to stack over consecutive images. The results of the analysis are also described in the next section.

\section{Results}
\label{sec:results}

After processing, 348 stacks were recorded as containing candidate difference signatures. Five categories were formed to describe the phenomena causing these candidate detections. The categories were formed by interrogating the pre-stacked difference images to determine if the difference signature was caused by a persistent signal over multiple images or due to bright signals in a small number of images. Candidates were categorised based on cadence, rotation angle, and the change in amplitude between consecutive images. In some cases, additional information such as the distance to an RSO and the angular speed were calculated to definitively categorise a detection. The stacking and classification was able to be processed interactively. Excluding individual frames to see how the detection changed was useful to determine if there was low amplitude persistent signal without single image bright signals. The purpose of categorising the detections was to determine if a candidate was a confirmed RSO. The following results detail each of these five categories:

\subsection{Detections due to a single short duration signal}
Detections in this category contained a single short duration signal (a burst) that appeared in only one time step along the predicted trajectory of the RSO. 65\% of the candidates fell into this category, with 82\% in the false positive test. A simple $\chi$-squared test shows there is no statistical difference between these results, suggesting that the detections in this category are not due to RSOs.

In difference space, the burst is positive in frame $\Delta I_{n + 1}$ and negative in the subtracted $\Delta I_{n}$ frame. The predicted position of the RSO under test moves between frames, meaning that $\Delta I_{n + 1}$ and $\Delta I_{n}$ are slightly offset when stacked. This mimics the difference signature of an RSO and is why the ability to efficiently interrogate the pre-stacked images is very useful. The likely cause of these signals is the reflection of terrestrial radio emission off the ionised plasma in meteor trails \citep{2014ApJ...788L..26O,2018MNRAS.477.5167Z} and the inexact differencing of bright sources \citep{2020PASA...37...39T}.  These meteor reflections can then randomly align with the trajectory of an RSO.

The final $\sim1.8$ hours of the observation time was contaminated by lightning and this manifested as greatly increased noise on the Western horizon. There were three cases of this lightning causing the stack to be a candidate.

\subsection{Detections due to two short duration signals}
This category encompasses the presence of two temporally distinct signals contributing to the difference signature in the stacked image, as opposed to a single signal as described above. 4.6\% of the stacks fell into this category, with 4.5\% in the false positives test.  Again, this difference is not statistically significant. Therefore it is unlikely that these are RSO detections and are most likely to be chance coincidences of two meteor reflections along the RSO trajectory.

\subsection{Detection of aircraft}

Aircraft prove to be efficient reflectors of radio waves in the 98.4375 MHz channel as they are much larger than most RSOs and far closer to the EDA2. While detected RSOs are in the 400 - 1,500 km range, aircraft can be $<10$ km away from the EDA2. They produced many of the highest amplitude signals in the dataset. Peak amplitudes frequently exceeded 100,000 Jy/beam and this would completely contaminate an image with sidelobes. If the peak amplitude of an aircraft exceeded 8,000 to 10,000 Jy/beam, the image would likely exceed the maximum allowable 200 Jy/beam RMS threshold and be dropped by the processing flow. Aircraft were normally distinct from RSOs by their signal amplitude, angular speed, and size, but there were some cases closer to the horizon where further analysis was necessary. 10\% of candidate detections were due to aircraft.

Building an intuition of common flight paths, reflectivity changes, and the detection frequency of aircraft is important for other more sensitive astronomical studies where it is crucial to have high quality data free of RFI \citep{2019PASA...36...23T}. \texttt{Insight}'s picking tool and interactive Python processing were used to investigate this separately to the main RSO detection workflow. The picking tool enables a user to `hand place' an annotation on a position in the image and in this case it would find the closest maximum to the pick, which represented the signal from an aircraft. To better understand the frequency of flight paths, each aircraft was manually picked on each image. In this separate workflow which was designed solely for aircraft, the interactive Python processing used these picks to create a mask of pixels over 6$\sigma$, constrained to a radius of five pixels around the pick. Peak amplitudes, zenith angles, and the SNR of aircraft detections were then calculated interactively.

\begin{figure}
    \centering
    \includegraphics[width=0.5\textwidth]{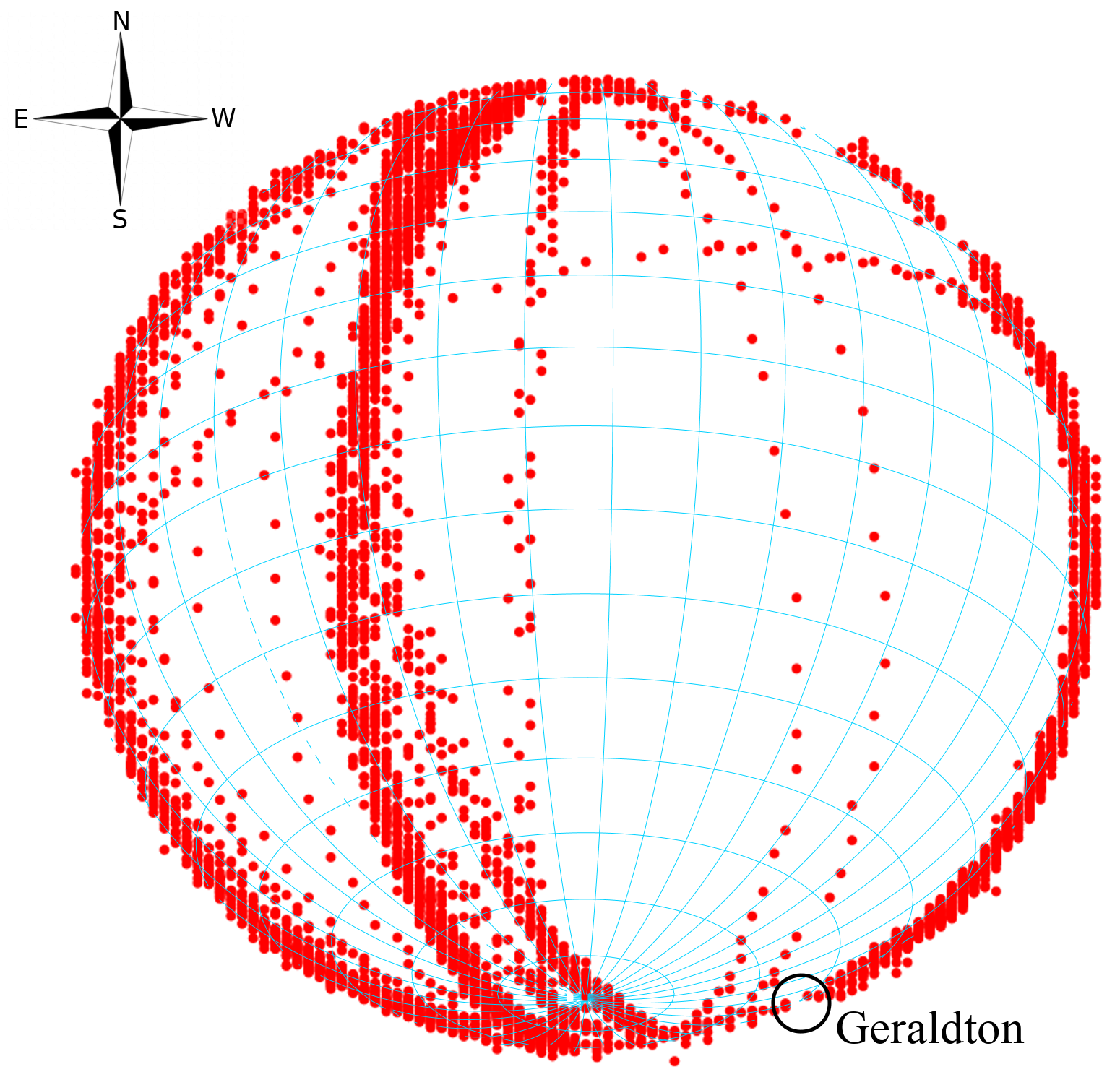}
    \caption{\label{fig:flight_paths} Locations of aircraft detected over the EDA2. The blue grid represents the visible area above the horizon. This image was taken from \texttt{Insight}'s image viewer.}
\end{figure}

Figure \ref{fig:flight_paths} shows the aircraft picks. This highlights the flight paths over the EDA2 during the three day recording period. 185 aircraft passes were detected in this time. As discussed in the previous work by \citet{2020PASA...37...39T}, many of these closer to the zenith will be aircraft flying to/from Perth, located South of the MRO. The less frequently traced flight paths are likely attributed to small aircraft headed to/from small mining towns which are numerous in northern and central Western Australia. 

The Perth International Airport had the vast majority of International flight traffic for Western Australia during the 2020/21 year\footnote{\url{https://www.bitre.gov.au/sites/default/files/documents/webairport_fy_1986-2021.xlsx}}. The largest domestic airport in close vicinity to the EDA2 is the Geraldton Domestic Airport, which had 441 inboard and 442 outboard aircraft in the 2020/21 year. This equates to a flux of $\sim2.4$ aircraft per day for a uniform distribution of flights.

For the North/South flying aircraft not along the horizon, 69 were flying North to South and only 20 from South to North. We assume these to be international flight routes. Monthly airport traffic data are available for Perth airport\footnote{\url{https://www.bitre.gov.au/sites/default/files/documents/webmonthlyairportjune2021.xlsx}}, recording 3,572 inbound and 3,581 outbound flights in February 2020. Assuming an even distribution across days of the month (which is probably less valid with the gradual reduction in flights due to the onset of the COVID-19 pandemic), there were 875 inbound international aircraft which equates to $\sim$30 international aircraft arriving in Perth each day. Assuming all 69 North to South aircraft in the dataset were international flights, then this equates to 23 per day. When acknowledging that flights are unlikely to be uniformly distributed across the month, the expected value is similar to the observed value. The remaining 7 per day could arrive at Perth from western or southern flight routes such as the Johannesburg to Perth route which would be out of range of the EDA2's sensing abilities.

If the utilisation of technologies such as the EDA2 are interesting for the purposes of detecting and tracking aircraft, the 256 antennas of the EDA2 are unnecessary for a cost-effective solution. To determine a suitable number of individual antennas needed for the detection of aircraft, the SNR of a subset of South flying (likely to Perth Airport) aircraft at an angle of $43\degree < \theta < 47\degree$ to the zenith was calculated. The average SNR of these 262 detections was $\sim$3,500. The signal was defined as the RMS of pixels above a 6$\sigma$ threshold and within 5 pixels of the centre of the aircraft. The noise was defined as the RMS of these same pixels 50 timesteps earlier. Assuming an SNR of 10 is adequate to sense an aircraft flying at a minimum angle of 45$\degree$ to the zenith, 8 - 16 EDA2 dipole antennas would be more than sufficient to detect the aircraft. This presents a cost-effective solution for passive aircraft detection from both a construction and maintenance perspective. 

We can detect aircraft at an altitude of 10,000 m to a distance of $\sim$100 km. This covers $\sim31,000$ km$^2$ of land, which is approximately the size of Belgium. If three of these sensors were used in an equilateral triangle formation, they could be positioned $\sim$170 km away from each other for complete overlapping coverage.

\subsection{Detection of RSOs}
\label{subsec:detections}

37 of the stacks (11\%) showed evidence of a detection, with none recorded in the false positives test. An additional three detections were made manually which were not flagged by the processing flow, taking the total number of detections to 40. These 40 detections are from 98,244 total passes of 6,110 RSOs over 75 hours. This marks an improvement of 34 detections compared to the previous work \citep{2020PASA...37...39T}, as well as a reclassification of FLOCK-3P-71 to BGUSAT. We categorise these into two types, A and B. Type A refers to when a detection is evident in the stack, but no obvious signals exist in any of the individual difference images at the predicted position of the RSO. Type B refers to detections where clear evidence of signal is returned at the predicted position of the RSO over multiple individual difference images and the stack. Details of these detections are described in Table \ref{tab:detections_pt1}.

\begin{table*}
\centering
\caption{Detected RSOs and their parameters}
\begin{adjustbox}{width=0.8\textwidth,center=\textwidth}
\begin{tabular}{cccccc} 
\hline\hline
Count & Satellite name & NORAD ID & RCS ($m^{2}$) & \begin{tabular}[c]{@{}c@{}}Closest distance of \\ pass ($km$)\end{tabular} & SNR   \\ 
\hline\hline
\multicolumn{6}{c}{The following detections were made on $31^{st}$ January 2020 (Observed 04:25:41– 23:59:59 UTC)} \\
\hline\hline
1              & TRITON 1        & 39,427    & 0.1 - 1.0    & 1,007   & 26.5    \\
\hline
2*             & ARIANE 40+ R/B  & 23,561    &  $>1.0$      & 813    & 6.4     \\
\hline
3              & BGUSAT          & 41,999    &  $<0.1$      & 537    & 28.9    \\
\hline
4              & YAOGAN 26       & 40,362    &  $>1.0$      & 513    & 11.7    \\
\hline
5*             & STARLINK 1044   & 44,749    &  $>1.0$      & 554    & 10.2    \\
\hline
6              & ISS (ZARYA)     & 25,544    &  $>1.0$      & 577    & 29.6    \\
\hline\hline
\multicolumn{6}{c}{The following detections were made on $1^{st}$ February 2020 (Observed 00:00:00 – 23:59:59 UTC)}\\
\hline\hline
7              & BGUSAT          & 41,999    &  $<0.1$      & 1,459   & 16.0    \\
\hline
8              & ISS (ZARYA)     & 25,544    &  $>1.0$      & 1,045   & 11.1    \\
\hline
9              & BGUSAT          & 41,999    &  $<0.1$      & 1,120   & 10.6    \\
\hline
10             & ISS (ZARYA)     & 25,544    &  $>1.0$      & 942    & 33.2    \\
\hline
11             & KONDOR E        & 40,353    &  $>1.0$      & 459    & 8.0     \\
\hline
12             & AQUA            & 27,424    &  $>1.0$      & 727    & 11.0    \\
\hline
13             & AURA            & 28,376    &  $>1.0$      & 725    & 12.2    \\
\hline
14             & ADEOS 2         & 27,597    &  $>1.0$      & 847    & 6.5     \\
\hline
15             & TRITON 1        & 39,427    & 0.1 - 1.0    & 779    & 19.9    \\
\hline
16*            & METEOR M1       & 35,865    &  $>1.0$      & 834    & 8.2     \\
\hline
17$^{\dagger}$ & SAOCOM 1-A      & 43,641    &  $>1.0$      & 639    & 12.9    \\
\hline
18             & HST             & 20,580    &  $>1.0$      & 585    & 8.4     \\
\hline
19             & COSMOS 1633     & 15,592    &  $>1.0$      & 484    & 6.2     \\
\hline
20             & HST             & 20,580    &  $>1.0$      & 546    & 20.8    \\
\hline
21             & MAX VALIER SAT  & 42,778    & 0.1 - 1.0    & 547    & 34.3    \\
\hline
22             & BGUSAT          & 41,999    &  $<0.1$      & 650    & 33.0    \\
\hline
23             & ISS (ZARYA)     & 25,544    &  $>1.0$      & 968    & 24.2    \\
\hline
24             & RISAT-2BR1      & 44,857    &  $>1.0$      & 577    & 9.0     \\
\hline
25             & ISS (ZARYA)     & 25,544    &  $>1.0$      & 1,019   & 11.2    \\
\hline
26$^{\dagger}$ & TRITON 1        & 39,427    & 0.1 - 1.0    & 981    & 24.0    \\
\hline\hline
\multicolumn{6}{c}{The following detections were made on $2^{nd}$ February 2020 (Observed 00:00:00 – 23:59:59 UTC)}\\
\hline\hline
27             & BGUSAT          & 41,999    &  $<0.1$      & 626     & 23.5   \\
\hline
28             & ISS (ZARYA)     & 25,544    &  $>1.0$      & 560     & 30.5   \\
\hline
29$^{\dagger}$ & COSMOS 1825     & 17,566    &  $>1.0$      & 537     & 11.0   \\
\hline
30*            & TITAN 4B R/B    & 28,647    &  $>1.0$      & 544     & 6.2    \\
\hline
31             & HST             & 20,580    &  $>1.0$      & 550     & 21.4   \\
\hline
32             & MAX VALIER SAT  & 42,778    & 0.1 - 1.0    & 864    & 28.6    \\
\hline
33             & BGUSAT          & 41,999    &  $<0.1$      & 1,155   & 25.0    \\
\hline
34             & ISS (ZARYA)     & 25,544    &  $>1.0$      & 1,427   & 10.4    \\
\hline
35             & ISS (ZARYA)     & 25,544    &  $>1.0$      & 663    & 33.8    \\
\hline
36             & TRITON 1        & 39,427    & 0.1 - 1.0    & 635    & 33.7    \\
\hline\hline
\multicolumn{6}{c}{The following detections were made on $3^{rd}$ February 2020 (Observed 00:00:00 – 07:17:43 UTC)}\\
\hline\hline
37             & MAX VALIER SAT  & 42,778    & 0.1 - 1.0    & 530    & 25.1    \\
\hline
38             & ISS (ZARYA)     & 25,544    &  $>1.0$      & 440    & 30.5    \\
\hline
39             & BGUSAT          & 41,999    &  $<0.1$      & 556    & 31.2    \\
\hline
40*            & SL-16 R/B       & 16,182    &  $>1.0$      & 1,003   & 7.8     \\
\hline\hline
\multicolumn{6}{l}{\parbox[t]{1.65\columnwidth}{Trajectories were propagated with TLE information from \texttt{space-track.org}. The website classifies the Radar Cross Section (RCS) into three categories: small ($<0.1$ m$^{2}$), medium (0.1 - 1.0 m$^{2}$), and large ($>1.0$ m$^{2}$). Detections labelled with * are type A and all others are type B. Detections labelled with $^{\dagger}$ were only found manually. These will be discussed further in Section \ref{subsec:detections}.}}
\end{tabular}
%\multicolumn{6}{l}{Continued...}      \\
\end{adjustbox}
\label{tab:detections_pt1}
\end{table*}

Five examples of type A detections were identified in the dataset. These are of interest because they represent a detection which would not have been possible by examination of the individual difference images alone. The weak signal returned from these detections is below the noise floor in the individual difference images, but when the rotations, shifts, and stacking have been applied, the SNR is increased. The individual difference images were manually inspected for any noise which could contribute to the difference signature that appeared in the stack. 

Three tests were conducted on these five examples to confirm they were detections.

In the first test, each half of the pass was separated and stacked to see if the difference signature was still visible in both. This was done interactively and individual frames or noisy sections could be removed to see how this affected the stack. Although this decreased the SNR, it was a useful tool to exclude candidates where noise contributing to the difference signature shape was not immediately obvious.

The second test was to visually inspect the stacks (and the corresponding difference images) of other passes of the RSO to see if there were missed detections. 

The final test was to research information on the physical characteristics of the RSO to determine if it was physically capable of causing a reflection. Valid information included pictures of the RSO, estimation of Radar Cross Section (RCS), physical size, any long antennas protruding from the RSO, identification as debris (more likely to be tumbling) and whether it was still commissioned. This information is only taken lightly into consideration as the reflective properties of an object are different in the radio band than the visual band. Visual light is on the order of nanometres ($10^{-9}$) while the wavelength of the 98.4375 MHz channel of this dataset is equivalent to $\sim$3 m. In theory, small RSOs such as cubesats should not be resolved by FM band radio waves unless they have large solar panels or an antenna which is capable of reflecting the long radio wavelengths. 

The closest approach of each RSO to the EDA2 for each detection is shown in Table \ref{tab:detections_pt1}. Type A detection SL-16 R/B is at the threshold of the EDA2's sensitivity at the closest point of its pass. SL-16 R/B has been previously detected by the MWA at a range of 863 - 873 km and peak flux density of 137.9 Jy/Beam \citep{2020PASA...37...52P}. The EDA2 has $1/8^{th}$ the sensitivity of the MWA, therefore a similar observation by the EDA2 would have 8 times the noise with the signal below the noise floor. Stacking allows the noise to be reduced, thus realising a type A detection.

A total of 32 type B detections were made in this analysis and are also described in Table \ref{tab:detections_pt1}.

All three detections of the ISS from the previous work were less than 600 km from the sensor at the closest approach of the pass \citep{2020PASA...37...39T}. The new algorithms in this work allow the detection of the ISS beyond a range of 1,400 km. We make all detections of the ISS when it passes less than 1,100 km from the EDA2 as shown in Table \ref{tab:iss_dist}. There is a final detection at 1,427 km and notably two additional passes at 1,381 and 1,406 km that are not detected. In these two passes, noise obfuscates any faint signal which may have been returned. The orientation angle of the ISS may also be less favourable to reflect any signal during these passes and explain why no detection is made in either of these.

\begin{table}%{R}{0.5\textwidth}
\centering
\caption{The closest distance between the EDA2 and the ISS for each pass during the survey.}
\begin{tabular}{ccc} 
\hline\hline
ISS pass & Closest approach (km) & Detected \\ 
\hline\hline
1         & 577   & Yes   \\
\hline
2        & 1,381   & No   \\
\hline
3        & 2,319   & No   \\
\hline
4         & 1,045  & Yes  \\
\hline
5         & 942   & Yes   \\
\hline
6         & 968   & Yes   \\
\hline
7         & 1,019  & Yes  \\
\hline
8        & 2,301   & No   \\
\hline
9        & 1,406   & No   \\
\hline
10        & 560   & Yes   \\
\hline
11        & 1,427  & Yes  \\
\hline
12        & 663   & Yes   \\
\hline
13       & 2,032   & No   \\
\hline
14       & 1,746   & No   \\
\hline
15        & 440   & Yes   \\
\hline
16       & 1,882   & No   \\
\hline\hline
\end{tabular}
\label{tab:iss_dist}
\end{table}

For an RSO equidistant from both the FM radio transmitter and the sensor, the received power from the reflection decreases approximately as a function of $\sim \frac{1}{d^{4}}$, where $d$ is the distance between the RSO and the source/sensor. For example, assuming the distance from the source to the reflector is equal to the distance from the receiver to the reflector and that the reflection surface of the ISS is constant, the difference in returned power between the furthest detection of the ISS (1,427 km) and the next closest pass (1,746 km) is a factor of $\sim2.2$.

The three weak detections of the Hubble Space Telescope (HST) are all less than 600 km away at the closest approach of the pass. This demonstrates how decreasing the physical size of the RSO severely diminishes its ability to reflect signal. The RCS of the ISS is 402 m$^{2}$ and the HST 29 m$^{2}$ \footnote{www.n2yo.com}. We could begin to characterise the reflective properties of individual RSOs in the 98.4375 MHz band with a larger sample size of detections. The ISS also has elongated, rectangular solar panels which are likely why it reflects the FM band radio signal so efficiently.

Of interest are detections of BGUSAT out to 1,459 km with a SNR of 16.0. In the previous work, it was shown that BGUSAT is likely transmitting out-of-band \citep{2020PASA...37...39T}. The received power of a directly transmitted beam instead follows a $\sim \frac{1}{d^{2}}$ law. Depending on the output power and directionality of the transmitted beam, there is the potential to detect this at a larger distance than a typical reflector. We are unable to detect the ISS past 1,427 km, yet we are able to detect the cubesat BGUSAT out to 1,459 km with a strong SNR, supporting the argument that BGUSAT is transmitting out-of-band. MAX VALIER SAT and TRITON 1 are both detected out to 864 and 1,007 km, respectively. Instead of their amplitude following the normal decay curve as an RSO passes over the sky, they sporadically produce signals at seemingly random intervals across the path. Seeing as they are also both small cubesats, we believe that they too are transmitting out of band.

Three detections were made which did not pass the processing flow parameterisation (detailed in Section \ref{sec:rotation}) and are marked with a $^{\dagger}$ in Table \ref{tab:detections_pt1}. These were of SAOCOM 1-A, TRITON 1, and COSMOS 1825. The SAOCOM 1-A and COSMOS 1825 detections were found while interactively interrogating the data for the purposes of the tests described above. The TRITON 1 detection was found when conducting a manual analysis of all other passes of TRITON 1 after it had been detected by the processing flow. This was standard for any detection made, as a test of the efficacy of the processing flow. In all cases, one of the selection criteria was not met, just falling beyond the thresholds set for the workflow. This illustrates that noise can warp the difference signature, meaning the fitted parameters of the Gaussian in Equation \ref{gaussian} do not accurately reflect the true shape of the signal and as a result the stack is rejected. Choosing more lenient selection criteria will lead to more false positives, so the parameterisation is a balance between confidence of detection and maximising the number of detections relative to false positives.

\begin{figure*}
    \centering
    \includegraphics[width=.9\textwidth]{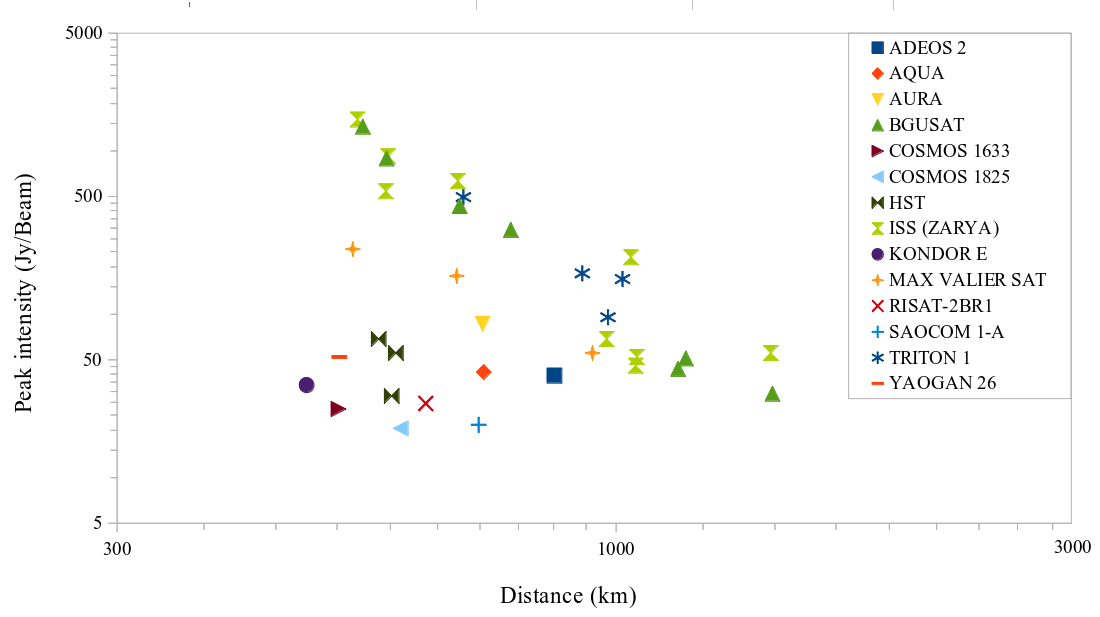}
    \caption{\label{fig:dvpa} The distance between the EDA2 and the RSO for type B and manual detections (N = 35) at the time of peak intensity of the pass. Both axes are log scale and have been truncated away from 0.}
\end{figure*}

Figure \ref{fig:dvpa} shows the relationship between the distance to an RSO and its peak intensity for type B and manual detections. The variety of RCSs between RSOs means there is no general predicted or observed overall trend. The beam response of dipoles in the EDA2, the orientation of the reflective surface of the RSO, the character of the transmitted radio signal from the source, the distance between the RSO and the source or receiver, noise, and background sources of RFI can all affect the measured reflected signal from an RSO at any point in time. This is why multiple observations of an RSO are not expected to fit a perfect power law. It is of interest to note that detections of BGUSAT seem to obey a power law, although more observations are needed to definitively quantify a relationship between peak intensity and distance.

\begin{figure}
    \centering
    \includegraphics[width=.46\textwidth]{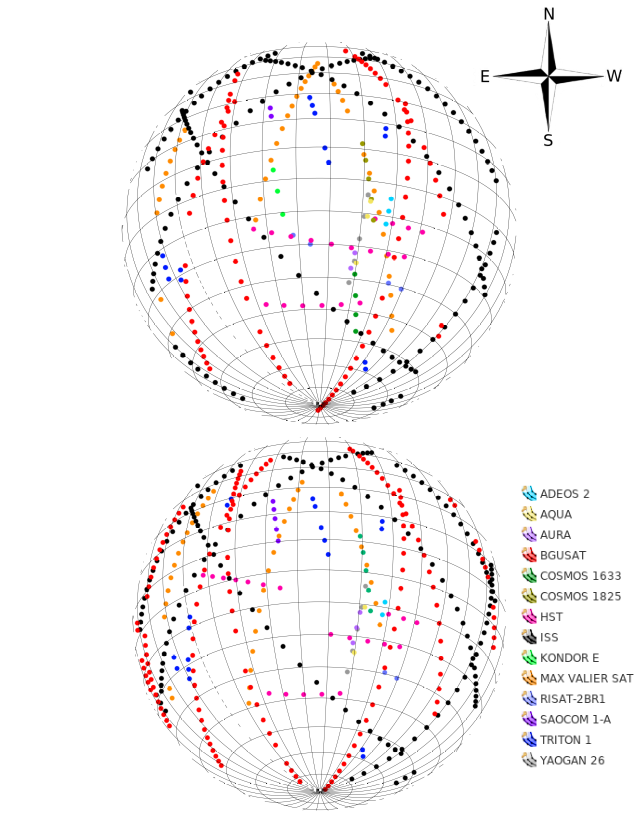}
    \caption{\label{fig:dmap} The distribution maps of type B and manual RSO detections (N = 35) for the XX (left) and YY (right) polarisations. This image was taken from \texttt{Insight}'s image viewer.}
\end{figure}

The beam power patterns shown in Figure \ref{fig:beams} have an effect on the spatial distribution of detections. Figure \ref{fig:dmap} shows the difference between the two polarisations. For this analysis, the two polarisations were kept separate and not stacked together, with the difference imaging being the only processing done to the data. This was to see the contribution of each polarisation to each of the detections. There are more detections in the North/South extremities for the XX polarisation and more in the East/West extremities for the YY polarisation. The effect is especially pronounced for detections of BGUSAT, showing a polarised signal favouring the YY polarisation. Building detailed information on the character and behaviour of observed signal for each RSO is useful information for further research and to assess the performance of the EDA2 compared to other interferometric arrays.

\begin{figure*}
    \centering
    \includegraphics[width=.8\textwidth]{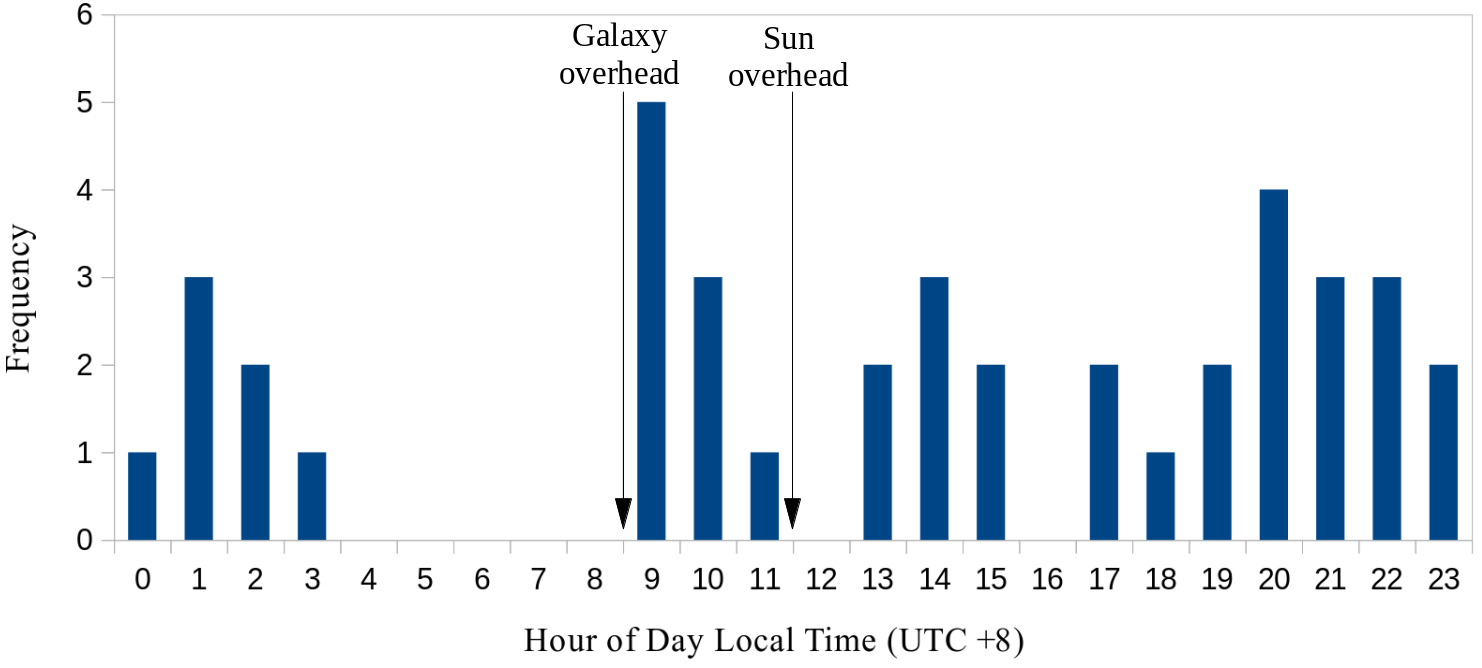}
    \caption{\label{fig:dph} Detections (N = 40) per binned hour of the day. Hour 0 corresponds to 12 AM midnight local time. The two arrows point to the nearest hour where the sun and Milky Way centre are directly overhead.}
\end{figure*}

The distribution of RSO detections as a function of time of day in Figure \ref{fig:dph} shows no obvious trend between the two variables. Generally, the RMS noise in an image is lowest 12 hours after the Milky Way is overhead. This equates to roughly 9 PM local time at the epoch of observation. There are 19 detections in daylight hours (6 AM to 7 PM) compared to 21 while the sun is down. This shows the critical importance of not discarding images where the sun is close to the zenith. A simple two-tailed Kolmogorov–Smirnov test shows that the distribution of detections is significantly different to a uniform distribution (p = 0.0122). The 5 hours between 4 AM - 9 AM local time contain no detections over any of the three days, but the sample size is not large enough to draw conclusions from this. The maximum amplitude at the location of Geraldton does not show any daily periodic oscillation which would imply that the output power of the FM transmitter decreases over these hours. 

\subsection{Detection of a different RSO}
33 stacks incorrectly identified RSOs to the wrong NORAD ID. There were two causes for this misidentification. The first was where the paths of two RSOs intersect and the same RSO is detected in two stacks under different NORAD IDs. Although the detection looks viable, when inspecting the difference images which contribute to the stack, it is trivial to discard the detection to the incorrect NORAD ID.

The second is where two RSOs have extremely similar orbital parameters for the entire pass. SOYUZ MS-13, SOYUZ MS-15, and PROGRESS MS-13 were all detected as they were docked to the ISS at the time of this survey. This meant that their predicted positions were very similar to that of the ISS and hence were flagged as detections. Cygnus NG-12 was also detected as it had unberthed from the ISS 4 hours prior to detection and the predicted position was still close enough to the ISS to be detected by the pipeline. Therefore, these were resolved with knowledge of the functions of the spacecraft and future incorrect detections such as these could be mitigated by excluding corresponding TLEs for the period in which spacecraft are docked to the ISS.

There were two stacks which were more difficult to classify, as the two RSOs were not docked to one another. FLOCK-3P-5 and BGUSAT were launched on the same payload on 2017-02-15 and have very similar orbits. To correctly identify which of the RSOs were responsible for the observed signal, other passes of the two RSOs had to be interrogated where it became evident that the signal was due to BGUSAT.

\section{Discussion}
\label{sec:discussion}

This study complements and significantly extends the work by \citet{2020PASA...37...39T} in temporally and spatially resolving RFI at the MRO to determine its origin. The selection of a singular frequency band which overlaps local FM radio broadcast allows the passive detection of backscatter from RSOs, meteors, and aircraft. Characterising these signals is important for predicting the effects of RFI on sensitive radio astronomy experiments and for the detection and identification of RSOs. 

\texttt{Insight}'s intuitive approach to building workflows was valuable in the design phase of the workflow. The hands-on manner of interrogating data from an early stage to identify anomalies, and the interactive parameter testing resulted in a highly optimised, scalable workflow which can be run on similar datasets acquired in the future.

A reproducible, production-ready workflow was built and run, detecting 37 individual passes of RSOs with an additional three manual detections, to in total identify 19 unique RSOs. Only seven passes from three unique RSOs were detected by Tingay et al. (Tingay et al. misidentified BGUSAT as FLOCK-3P-71, which has been resolved in the current work). For RSOs where backscatter of terrestrial FM radio is detected, this workflow was able to make detections of the ISS out to a distance of 1,427 km from the sensor (at the closest distance of the pass). This is an improvement from 577 km in the previous work \citep{2020PASA...37...39T} and implies a range of 1,500 km for the detection of non-transmitting RSOs by the EDA2, as the ISS was by far the largest in orbit at the time. Two passes of the ISS were within this range and not detected at a closest distance of the pass of 1,381 and 1,406 km. Upon scrutiny, the sidelobes created by both spurious noise and direct transmission from Geraldton increase the RMS of the images contributing to the stack and prevent these faint signals from being detected. This highlights the importance of RFI being minimised and shows that even though the Murchison region is classified as a Radio Quiet Zone (RQZ), that anthropogenic FM band radio signals still decrease the quality of this research.

A worrying class of these detections are cubesats transmitting out of band. Multiple high amplitude detections of BGUSAT with peak amplitudes exceeding 1,000 Jy/Beam were made. TRITON 1 and MAX VALIER SAT were also observed to be transmitting out of their allocated bands. The work by \citet{soko_eda2} also found BGUSAT, TRITON 1, and MAX VALIER SAT to be transmitting out of range at higher frequencies (and outside the allowed downlink frequencies for RSOs). Additional studies have identified UKube-1 and DUCHIFAT-1 to be exhibiting similar behaviour \cite{10.1093-mnras-sty930, 2020a}. This growing list of transmitting RSOs further contaminates all RQZs. More observations are needed to probe the character of these transmissions to determine if the transceivers are faulty, or if there is another reason for the out-of-band transmission.  Studies such as this, \citet{2020PASA...37...39T}, \citet{soko_eda2}, \citet{10.1093-mnras-sty930}, and \citet{2020a} demonstrate that the broadband nature of these signals is destructive to a range of radio astronomy research and can hopefully inform future discussions on which electronics are the most reliable for cheap satellites. With more data over different frequency bands we can determine how these observed signals match up to ITU regulations which is complex and out of scope for this paper.

The detection of a single STARLINK satellite does raise concerns about their visibility in radio astronomy research. Reasons that we have only detected one, when 117 STARLINK satellites satisfied the TLE requirements, could be because of variation in the transmitter power pattern, the variable radar cross section of the object, and the fact that catalogued RCS values are mainly at high frequency instead of low frequency. It is likely that only favourable transmitter, object, and receiver geometries will be detected and it will be worthwhile to test this over multiple frequency bands in the future. It is also possible that only STARLINK satellites that have faulty electronics and are transmitting out of band are detected. \citet{2020PASA...37...52P} encountered a similar scenario in their work.

Rather than characterising the general transient environment of the Murchison region, which previous papers have focused on, this paper looks more specifically at anthropogenic sources of RFI to inform future work on how these affect radio astronomy research. This research has shown that a near expected number of international aircraft are detected in the dataset. \citet{soko_eda2} suggests that the parallax effect of two nearby arrays can be used to generate a 3D flight path of these aircraft, demonstrating the benefit of using radio interferometry as an aircraft traffic monitoring tool. \citet{soko_eda2} also estimates an expected detection threshold for BGUSAT based on a radar theory and concludes that the observed amplitudes of BGUSAT in the dataset are far more likely caused by ``line-of-sight propagation from a low power transmitter ($\leq1 W$) with a small fraction of out-of-band `spill-over' over a wide frequency band". Their analysis of this was in the 159.375 MHz band while this analysis also supports this conclusion in the 98.4375 MHz band.

The detections outlined in Table \ref{tab:detections_pt1} probe the limit of the range of the EDA2. With the acquisition of more data by the EDA2, sensitivity limits can be built for each RSO and additional detections will be useful for determining the range at which the EDA2 can detect them.

This study makes 40 individual detections of RSOs compared to the 7 from the previous study \citep{2020PASA...37...39T}, which marks an improvement of almost 6 times. In Section \ref{sec:rotation}, it is shown that the minimum expected SNR improvement is $\sqrt{48} \approx 7$. It is promising to see that this work achieves approximately the minimum expected improvement and future work will aim to push this even higher.

A relevant feature of \texttt{Insight} is its ability to ingest and provide QC tools for the live acquisition of data. For example, \texttt{Insight} is currently used in a production environment for the QC of data as they are acquired by a seismic vessel. As each snapshot of the subsurface is acquired, \texttt{Insight} monitors a directory and passes the raw data through a series of pre-configured workflows and scripts to visualisations from which a user can perform live QC. Instant feedback of errors in the acquisition to the crew on the vessel avoids extended equipment malfunctions which may affect the quality of data. This workflow produces a series of incremental outputs which are shipped as QC products with the raw data. 

The workflow described in this study has the potential to be run live as data are acquired by an interferometric array. This would enable catalogue maintenance of RSO trajectories and for highlighting when aircraft are present over the array. In anticipation of SKA era processing, the data from a single SKA station (likely to contain 256 dipoles similar to the EDA2 configuration) can be processed live through a real-time pipeline. Detection and characterisation information for transient RFI can be included in the flags for acquired data, to warn that the data may contain RFI. For example, if future work identifies which frequency bands BGUSAT is wrongly transmitting over and out to which distance it is typically detected, this information can be included in the flags. This can be precomputed information and would not affect the real-time acquisition of SKA scale operations.

The potential for `small scale' passive radio-interferometry becomes apparent for surveillance purposes. If the goal is to detect aircraft (including drones) in Western Australia, the most cost-effective architecture would be 8 - 16 EDA2 style dipole antennas, which we predict would have a range of $\sim$100 km for aircraft flying at 10,000 m altitude. This provides a field of view over 31,000 km$^{2}$ or equivalent to approximately the size of Belgium. When stealth is critical, this passive option for surveillance is of great relevance. If the detection of any aircraft is prioritised over stealth, this option provides another dimension to pre-existing detection techniques such as higher frequency radar, optical, and infra-red band sensing. 

To enable the workflow in this study to operate in a real-time environment, there are some modifications which will need to be made. False positives identified by the workflow need to be reduced as ideally no human intervention would be required to categorise a detection (currently 11\% of stacks passing through the workflow are confirmed detections). Future efforts can be focused on subtracting static models of known astronomical radio emission before differencing images, to reduce the effect of differencing residuals. Modelling the expected response from a reflection as a function of the EDA2 beam response and the RSO position on the sky could increase the SNR of detections by ensuring the shape of the returned signal would be consistent when stacking. A more attractive method which may be useful for real time RSO catalogue maintenance would be to use artificial intelligence to make detections. To execute this additional study, a significant observational period would need to be acquired to train a network. Once detections have been made, TLEs can be computed over single passes and their accuracy improved over multiple passes \citep{9559621}. We will also aim to detect RSOs without prior knowledge of their expected positions, enabling the detection of unknown and uncatalogued RSOs.

Additional observations have been made by the EDA2 and AAVS2, similar to those described in \citet{soko_eda2}, which will be used to test and improve this processing flow. We hope to complement this work with an in-depth multi-frequency investigation of the passive detection of RSOs.

\section{Conclusion}
\label{sec:conclusion}

The trajectory of \texttt{Insight}'s development suggests strong overlap with the demand for scalable and flexible processing and visualisation in the era of SKA data processing. This proof of concept demonstrates \texttt{Insight}'s ability to process a complex radio-astronomy dataset and provide the user with the means to make intuitive processing decisions in an interactive environment. This intimate interaction with the dataset, through the use of the software, allowed parameter optimisation and process development which were far more tailored than existing methods and ultimately able to detect RSOs at a further distance than the previous work \citep{2020PASA...37...39T}.  Although this was a small-scale dataset, it is highly complex, containing a range of different signal types and using \texttt{Insight} was highly beneficial in qualitatively categorising these signals through statistical methods for the purpose of SSA. The workflow can be extended to much larger scale automated processing in the \texttt{Insight} framework for general purpose SSA or for identifying specific causes of RFI such as aircraft or RSOs.

\section{Acknowledgements}
\label{sec:acknowledgements}

Data processing was run at the DUG Technology HPC centre in Perth, Western Australia. \texttt{Insight} software has been developed and is maintained by DUG Technology whom Dylan Grigg is an employee of. This work utilised \texttt{MIRIAD} \citep{1995ASPC...77..433S} for the preprocessing and the following Python modules: \texttt{Skyfield} \citep{2019ascl.soft07024R}, \texttt{Astropy} \citep{2018AJ....156..123A}, and \texttt{Numpy} \citep{numpy} and we wish to acknowledge the developers for their work. We acknowledge the Wajarri Yamatji people as the traditional custodians of the Murchison Radio Observatory site. \citep{5164979}

\bibliographystyle{model2-names} 
\bibliography{cas-refs}

\end{document}